\documentclass[preprint]{aastex}
\newcommand{\h} {H{\sc i}}
\newcommand{\hp} {H{\sc ii}}
\newcommand{\he} {He\,{\sc i}}
\newcommand{\hep} {He\,{\sc ii}}
\newcommand{\hepp} {He\,{\sc iii}}
\newcommand{\tot}{{\rm\scriptscriptstyle {tot}} }
\newcommand{\gas}{{\rm\scriptscriptstyle {gas}} }
\newcommand{\DM}{{\rm\scriptscriptstyle {DM}} }
\slugcomment{RESCEU-33/99\quad UTAP-344/99}
\shorttitle{Cosmological SPH simulations}
\shortauthors{Yoshikawa, Jing \& Suto}
\begin{document}

\title{Cosmological SPH simulations with four million particles:\\
statistical properties of X-ray clusters in a low-density universe}

\author{Kohji Yoshikawa}
\affil{Department of Astronomy, Kyoto University, Kyoto 606-8502,
Japan.}
\email{kohji@kusastro.kyoto-u.ac.jp}

\author{Y.P. Jing and Yasushi Suto}
\affil{Department of Physics and Research Center for
    the Early Universe (RESCEU) \\ School of Science, University of
    Tokyo, Tokyo 113-0033, Japan.}
\email{jing@utap.phys.s.u-tokyo.ac.jp,suto@phys.s.u-tokyo.ac.jp}

\received{1999 September 14}
\accepted{1999 November ??}

\begin{abstract}
  We present results from a series of cosmological SPH (smoothed
  particle hydrodynamics) simulations coupled with the P$^3$M
  (Particle--Particle--Particle--Mesh) solver for the gravitational
  force. The simulations are designed to predict the statistical
  properties of X-ray clusters of galaxies as well as to study the
  formation of galaxies. We have seven simulation runs with different
  assumptions on the thermal state of the intracluster gas.  Following
  the recent work by Pearce et al., we modify our SPH algorithm so as
  to phenomenologically incorporate the galaxy formation by decoupling
  the cooled gas particles from the hot gas particles.  All the
  simulations employ $128^3$ particles both for dark matter and for
  gas components, and thus constitute the largest systematic catalogues
  of simulated clusters in the SPH method performed so far. These
  enable us to compare the analytical predictions on statistical
  properties of X-ray clusters against our direct simulation results
  in an unbiased manner.  We find that the luminosities of the
  simulated clusters are quite sensitive to the thermal history and
  also to the numerical resolution of the simulations, and thus are
  not reliable. On the other hand, the mass -- temperature relation
  for the simulated clusters is fairly insensitive to the assumptions
  of the thermal state of the intracluster gas, robust against the
  numerical resolution, and in fact agrees well with the analytic
  prediction. Therefore the prediction for the X-ray temperature
  function of clusters on the basis of the Press -- Schechter mass
  function and the virial equilibrium is fairly reliable.
\end{abstract}


\section{Introduction}

Clusters of galaxies have been widely used as probes to extract the
cosmological information. In particular, the cluster abundances
including the X-ray temperature function (XTF), X-ray luminosity
function (XLF) and number counts turn out to put strong constraints on
the cosmological density parameter $\Omega_0$ and the fluctuation
amplitude $\sigma_8$
\citep{Henry1991,White1993,Jing1994,Viana1996,Eke1996,Kitayama1996,Kitayama1997,Kitayama1998}.
Since the theoretical predictions for those purposes are usually based
on the Press -- Schechter mass function and the simple virial
equilibrium model for the X-ray clusters, the reliability of the
resulted constraints is heavily dependent on the validity of those
assumptions in the observed X-ray clusters. In fact there are numerous
realistic physical processes which would somehow invalidate the
simplifying assumptions employed in the above procedure; one-to-one
correspondence between a virialized halo and an X-ray cluster,
non-sphericity, substructure and merging, galaxy formation, radiative
cooling, heating by the UV background and the supernova energy
injection, etc.

While these can be addressed by hydrodynamical simulations in principle,
the relevant simulations are quite demanding. This is why most previous
studies simply check the Press -- Schechter {\it mass} function against
the purely $N$-body simulations in cosmological volume
\citep{Suto1993,Ueda1993,Lacey1994}, or focus on hydrodynamical
simulations of individual clusters in a constrained volume
\citep{Eke1998,Suginohara1998,Yoshikawa1998}; most previous
hydrodynamical simulations of clusters in cosmological volume, on the
other hand, lacked the numerical resolution to address the above
question, and/or ignored the effect of radiative cooling
\citep{Evrard1990,Cen1992,Cen1994,Kang1994,Bryan1998,Frenk1999} except
for a few latest state-of-the-art simulations
\citep{Pearce1999a,Pearce1999b,Cen1999}.

In this paper, we present results from a series of cosmological SPH
(smoothed particle hydrodynamics) simulations in cosmological volume
with sufficiently high resolution.  They enable us to compute the
statistical properties of simulated clusters in an unbiased manner, and
thus to directly address the validity of the widely used model
predictions for the mass -- temperature relation and the resulting XTFs.

\bigskip

\section{Simulations}

Our simulation code is based on the P$^{3}$M--SPH algorithm, and the
P$^{3}$M part of the code has been extensively used in our previous
$N$-body simulations \citep{Jing1994,Jing1998}. All the present runs
employ $N_{\rm DM}= 128^3$ dark matter particles and the same number
of gas particles. We use the spline (S2) softening function for
gravitational softening \citep{Hockney1981}, and the softening length
$\epsilon_{\rm grav}$ is set to $L_{\rm box}/(10N_{\rm DM}^{1/3})$,
where $L_{\rm box}$ is the comoving size of the simulation box.  
We set the minimum of
SPH smoothing length as $h_{\rm min}=\epsilon_{\rm grav}/4$, and adopt
the ideal gas equation of state with an adiabatic index of $5/3$.

We consider a spatially--flat low--density CDM (cold dark matter)
universe with the mean mass density parameter $\Omega_0=0.3$,
cosmological constant $\lambda_0=0.7$, and the Hubble constant in units
of 100 km s$^{-1}$ Mpc$^{-1}$, $h=0.7$.  The power-law index of the
primordial density fluctuation is set to $n=1$. We adopt the baryon
density parameter $\Omega_{\rm b}=0.015h^{-2}$ \citep{Copi1995} and the
rms density fluctuations on $8h^{-1}$ Mpc scale, $\sigma_8=1.0$,
according to the constraints from the {\it COBE} \/normalization and
cluster abundance \citep{Bunn1997,Kitayama1997}.

The initial conditions for particle positions and velocities at $z=25$
are generated using the COSMICS package \citep{Bertschinger1995}. We
prepare two different initial conditions for $L_{\rm
  box}=150h^{-1}$Mpc and $75h^{-1}$Mpc to examine the effects of the
numerical resolution. For the two different boxsizes, we consider
three cases for the ICM (intracluster medium) thermal evolution
(Table~\ref{tab:model}); L150A and L75A assume non-radiative evolution
of the ICM (hereafter non-radiative models).  L150C and L75C include
the effect of radiative cooling (pure cooling models). In L150UJ and
L75UJ (multiphase models), both radiative cooling and UV-background
heating are taken into account, and an additional modification of the
SPH algorithm is implemented in order to avoid the artificial
overcooling, following \citet{Pearce1999a,Pearce1999b}. In addition,
we have L150CJ which is identical to L150UJ but without the UV
background. While the L75C model is evolved until $z=0.5$, the
simulations of the remaining six models are computed up to $z=0$.

The radiative cooling and the UV-background heating take account of the
photoionization of {\h}, {\he} and {\hep}, the collisional ionization of
{\h}, {\he} and {\hep}, the recombination of {\hp}, {\hep} and {\hepp},
the dielectronic recombination of {\hep}, Compton cooling and the
thermal bremsstrahlung emission. The UV-background flux density $J(\nu)$
is assumed to be parameterized as
\begin{equation}
 J(\nu) = J_{21}(z)\left(\frac{\nu_{\rm L}}{\nu}\right)\times 10^{-21}
      \,[\rm erg/s/cm^2/sr/Hz],
\end{equation}
where $\nu_{\rm L}$ is the Lyman--limit frequency, and we adopt the
redshift evolution:
\begin{equation}
 \label{eqn:fluxdensity}
 J_{21}(z) = J_{21}^{0}\frac{(1+z)^4}{5+(1+z)^4}
\end{equation}
following \citet{Vedel1994}.  We use the semi-implicit scheme in
integrating the thermal energy equation (\citet{Katz1996}).

It is known that SPH simulations with the effect of radiative cooling
produce unacceptably dense cooled clumps.  While \citet{Suginohara1998}
ascribed this to some missing physical processes such as heat conduction
and heating from supernova explosions, \citet{Thacker1998} and
\citet{Pearce1999a} showed that this overcooling could be
due to an artificial overestimate of hot gas density convolved with the nearby
cold dense gas particles, and can be largely suppressed by simply
decoupling the cold gas ($T<10^4$K) from the hot component.  This
numerical treatment can be interpreted as a phenomenological
prescription for the galaxy formation.  Therefore following their
spirit, we adopt a slightly different approach --- using 
the Jeans criterion for the cooled gas:
\begin{equation}
 \label{eq:jeans}
 h > \frac{c_{s}}{\sqrt{\pi G \rho_\gas}} ,
\end{equation}
where $h$ is the smoothing length of gas particles, $\rho_\gas$ the
gas density, $c_s$ the sound speed, and $G$ the gravitational
constant.  In practice, however, we made sure that the above condition
is almost identical to the condition of $T<10^4$K adopted by
\citet{Pearce1999a}.  Nevertheless we prefer using the Jeans criterion
because it is rephrased in terms of the physical process. Apart from
the fact that such cooled gas particles are ignored when computing the
SPH density of hot gas particles, all the other SPH interactions are
left unchanged.

\bigskip

\section{Results}

\subsection{Cluster Identification and Radial Profiles}

At each epoch, we identify gravitationally bound objects using an
adaptive ``friend-of-friend'' algorithm \citep{Suto1992}, and select
objects with more than 200 dark matter and gas particles as clusters
of galaxies. Specifically we use the SPH gas density in assigning the
local linking length with the linking parameter of $0.5$. With this
parameter, the mass function of the selected clusters is nearly
identical to the conventional ``friend-of-friend'' with the fixed
linking length of $0.2$ times the mean particle separation.  The
virial mass $M$ for each cluster is defined as the total mass at the
virial radius within which the averaged mass density is $\simeq
18\pi^2 \Omega_0^{0.4}\bar{\rho}_c(z)$ predicted from non-linear
spherical collapse model, where $\bar{\rho}_c(z)$ is the critical
density of the universe at $z$ \citep{White1993,Kitayama1996}. The
X-ray luminosity is computed on the basis of the bolometric and band
limited thermal bremsstrahlung emissivity \citep{Rybicki1979}, which
ignores metal line emission. We also compute the mass weighted and
emission weighted temperatures, $T_X^{\rm m}$ and $T_{X}$, using
$2$--$10$ keV band emission. For models L150UJ and L75UJ, we do not
include contribution of those cold particles which satisfy the
criterion (\ref{eq:jeans}) in computing $T_X$, $T_X^{\rm m}$ and $L$,
since they are not supposed to remain as a gaseous component if an
appropriate star formation scheme is implemented in our simulation.

\begin{figure}[tbph]
 \begin{center}
  \includegraphics[height=15cm,keepaspectratio]{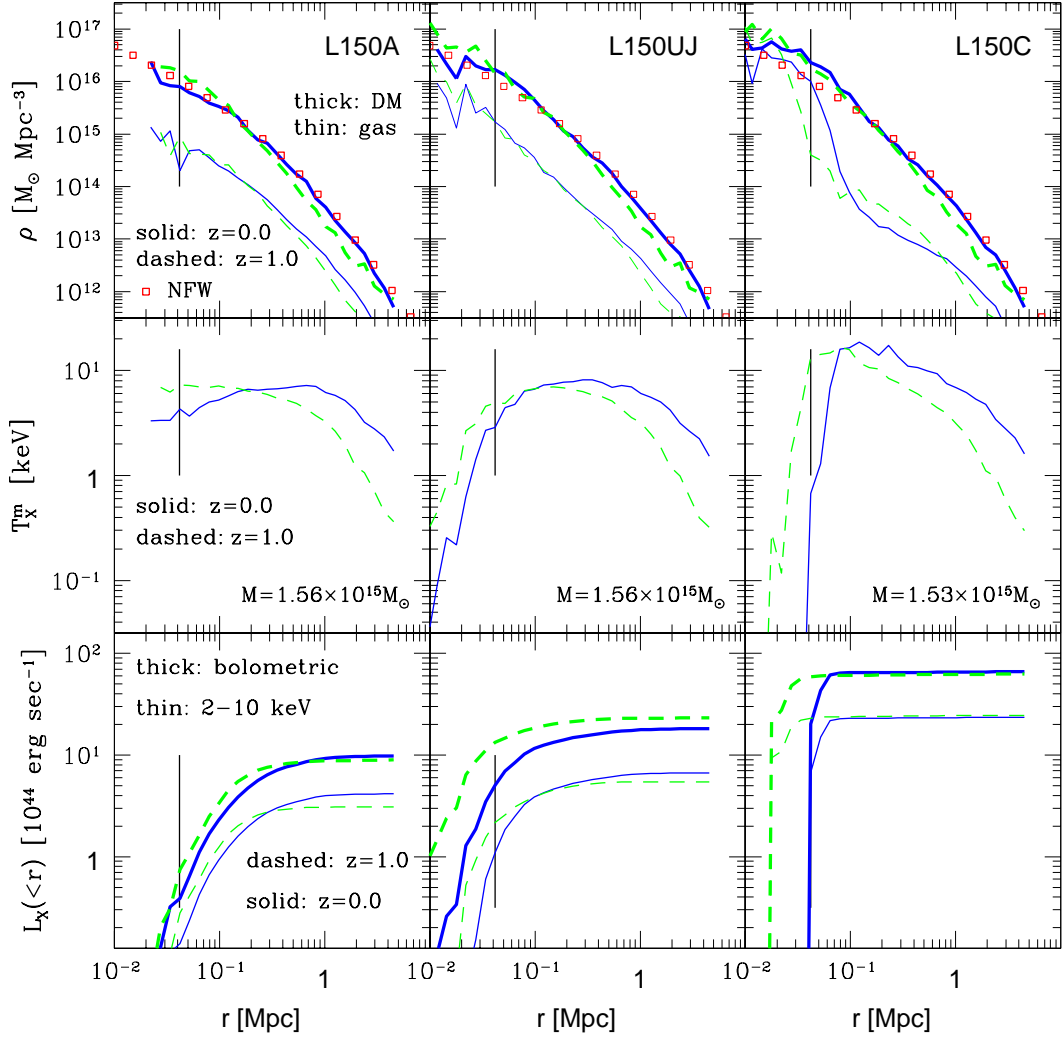}
 \end{center}
 \figcaption[profile.150.ps]{Spherically averaged profiles of dark
   matter and gas densities, ICM temperature and the integrated X-ray
   luminosity from the center, for a cluster at $z=0.0$ (solid lines)
   and $z=1.0$ (dashed lines) in models L150A, L150UJ and L150C.
   Vertical lines indicate our lower limit on the SPH smoothing
   length, $h_{\rm min}$. The viral mass of each cluster at $z=0.0$ is
   quoted in the middle panels. For reference, the universal density
   profile at $z=0$ \citep{NFW1997} corresponding to the virial mass of the
   cluster is plotted in open squares (top panels).
 \label{fig:profile150}}
\end{figure}

\begin{figure}[tbph]
 \begin{center}
  \includegraphics[height=15cm,keepaspectratio]{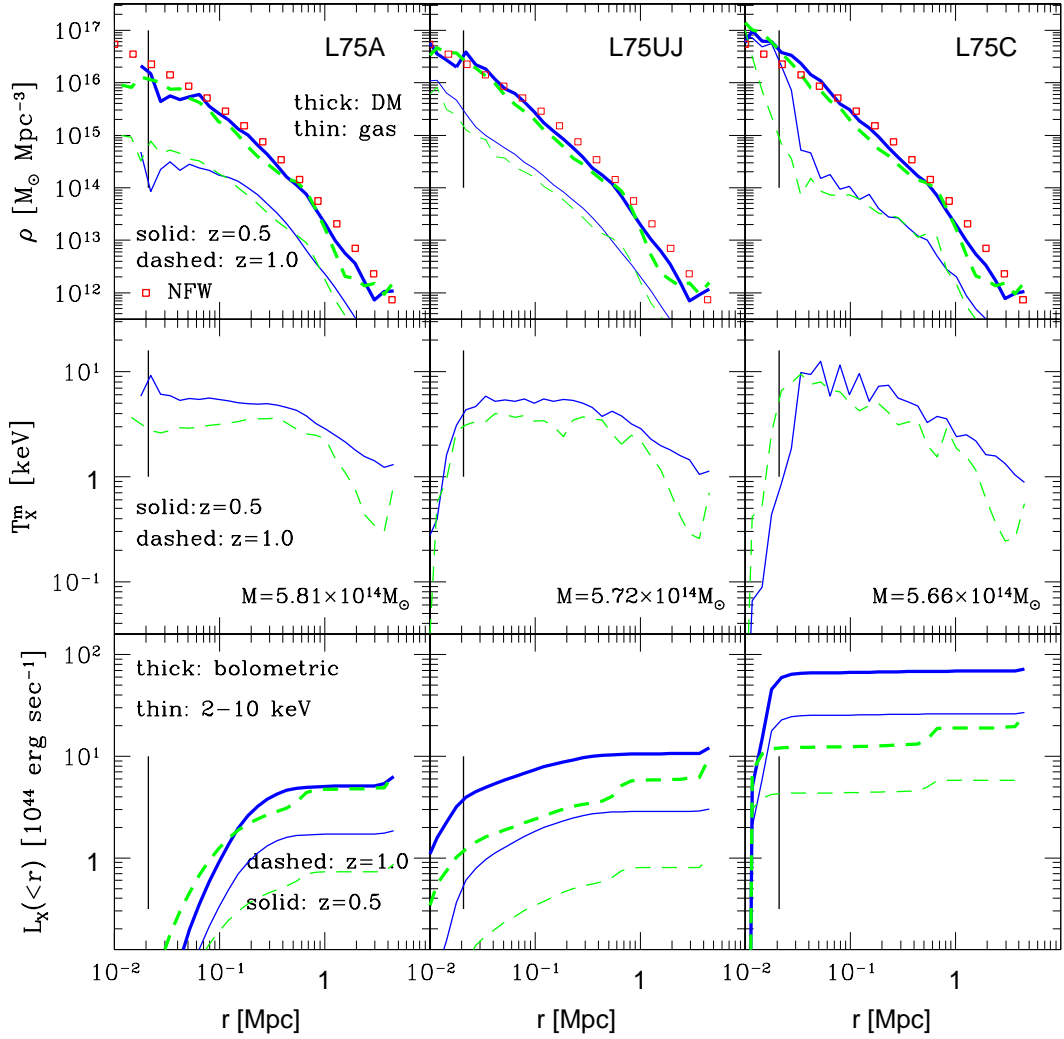}
 \end{center}
 \figcaption[profile.75.ps]{ Same as Figure \ref{fig:profile150} but
   for models L75A, L75UJ, and L75C at $z=0.5$ (solid lines) and
   $z=1.0$ (dashed lines). The virial mass of each cluster at $z=0.5$
   is quoted in the middle panels.\label{fig:profile75}}
\end{figure}

In Figure~\ref{fig:profile150}, we show the spherically averaged
profiles of dark matter and gas densities, ICM temperature, and
cumulative X-ray luminosity for the most massive cluster in models
L150A, L150UJ and L150C. Those for models L75A, L75UJ and L75C are
also depicted in Figure~\ref{fig:profile75}. The centers of clusters
are defined as the peaks of gas density. The profiles of dark matter
are fairly well-approximated by the model proposed by \citep{NFW1997}.
Due to an artificial cooling catastrophe, the pure-cooling model
(L150C and L75C) produces unacceptably high X-ray luminosity,
consistent with \citet{Suginohara1998}.  This artificial feature can
be largely removed by decoupling the cold gas from hot component, and
models L150UJ and L75UJ yield a fairly reasonable range of X-ray
luminosities.  Nevertheless we do not claim that the luminosities in
L150UJ and L75UJ are reliable; the current modified scheme is to be
interpreted as a tentative and phenomenological remedy at best and
should be replaced by more realistic scheme for galaxy formation.
\citet{Lewis99} have obtained a very similar result based on a more
sophisticated treatment of galaxy formation in their simulation. They
show that when both cooling and star formation are included in the
simulation, cluster luminosities increase only moderately (a factor 3)
compared to the non-radiative case; this amount of increase is
significantly less than that in the pure-cooling case. In addition, we
find that the values of the luminosities are still significantly
affected by the numerical resolution (\S \ref{subsec:tl}).

\begin{figure}[tbph]
 \begin{center}
  \includegraphics[width=15cm,keepaspectratio]{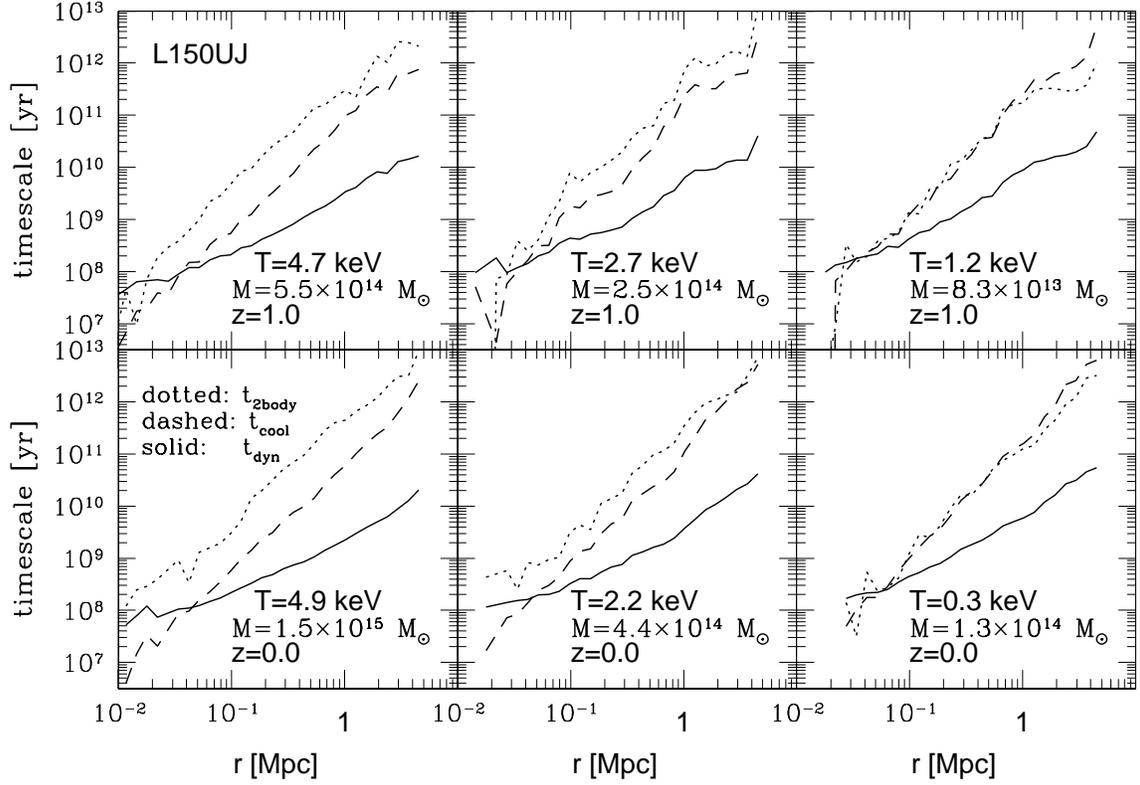}
  \end{center} \figcaption[timescale.150.ps]{Radial profiles of the
  local dynamical timescale, $t_{\rm dyn}$, the artificial two-body
  heating timescale $t_{\rm 2body}$ and the cooling timescale $t_{\rm
  cool}$ for three different clusters at $z=0.0$ (lower panels) and
  $z=1.0$ (upper panels) in model L150UJ. Their virial mass and
  temperature are indicated in each panel.\label{fig:timescale150}}
\end{figure}

\begin{figure}[tbph]
 \begin{center}
  \includegraphics[width=15cm,keepaspectratio]{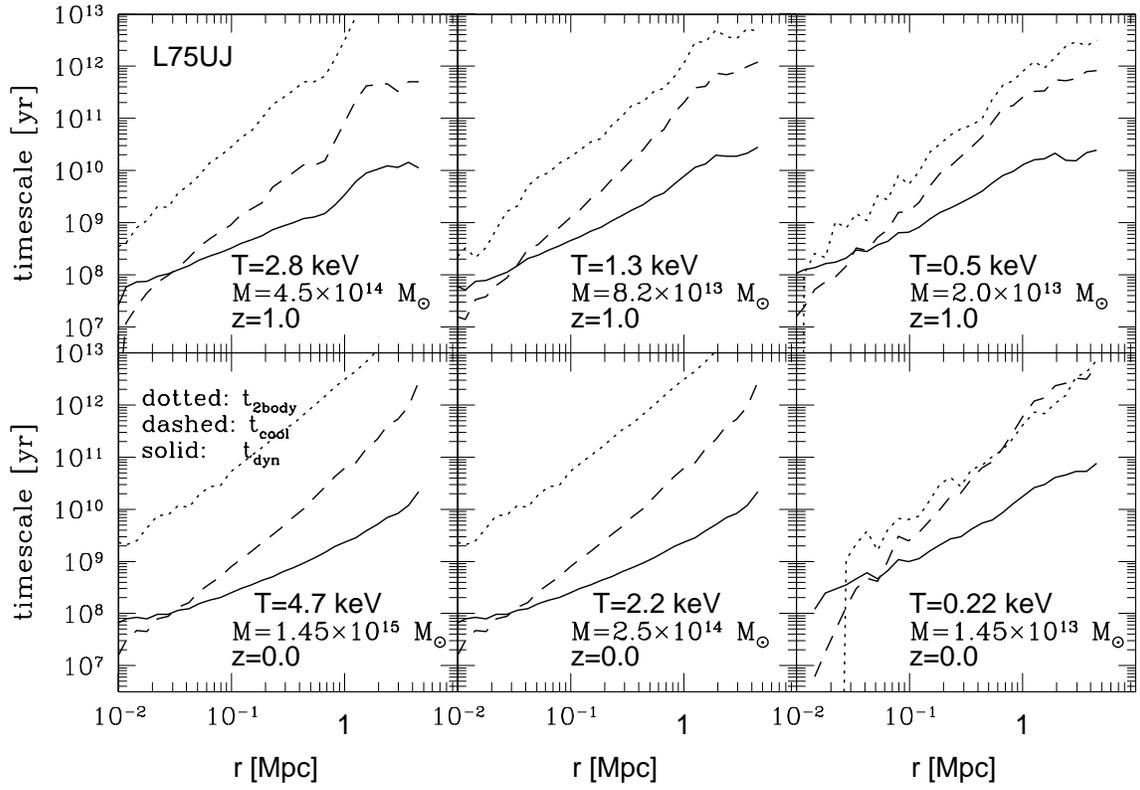}
 \end{center}
 \figcaption[timescale.75.ps]{Same as Figure~\ref{fig:timescale150} but
 for model L75UJ.\label{fig:timescale75}}
\end{figure}

In Figures~\ref{fig:timescale150} and \ref{fig:timescale75}, we show the
spherically averaged profiles of local dynamical timescale $t_{\rm
dyn}$, 2-body heating timescale $t_{\rm 2body}$ \citep{Steinmetz1997}
and cooling timescale $t_{\rm cool}$ for representative clusters with 3
different mass scale for models L150UJ and L75UJ, respectively. Each
timescale is defined as 
\begin{eqnarray}
 t_{\rm dyn}&=&\frac{1}{\sqrt{G\rho_\tot}} ,\\
 t_{\rm 2body}&=&\sqrt{\frac{27}{128\pi}}
  \frac{\sigma^3_{\rm 1D}}{G^2 M_{\DM}\rho_{\DM} \ln\Lambda} ,
\label{eq:2-body}\\
 t_{\rm cool}&=&\frac{3n_\gas k_{\rm B}T}{2\Lambda_{\rm cool}(n,T)} ,
\end{eqnarray}
where $G$ is the gravitational constant, $\rho_\tot$ the total mass
density , $\sigma_{\rm 1D}$ the 1-dimensional velocity dispersion of
dark matter particles, $\ln\Lambda$ the Coulomb logarithm, $M_{\DM}$
the mass of a dark matter particle, $\rho_{\DM}$ the mass density of
dark matter, $n_\gas$ the number density of gas, $k_{\rm B}$ the
Boltzmann constant and $\Lambda_{\rm cool}(n,T)$ the cooling function.
The expression for $t_{\rm 2body}$ is given in \citet{Steinmetz1997}
and we set the value of the Coulomb logarithm to $\ln\Lambda=5$ as a
nominal value. 

For the clusters with $M\gtrsim 2\times10^{14}\,M_{\odot}$ in the
models with $L_{\rm box}=150h^{-1}\mbox{Mpc}$ , $t_{\rm cool}$ is
shorter than $t_{\rm 2body}$, and hence the artificial 2-body heating
is not effective for these clusters. On the other hand, clusters with
$M\lesssim10^{14}M_{\odot}$ have the cooling timescale comparable with
the 2-body heating timescale. Thus, the relatively poor clusters with
$M\lesssim10^{14}M_{\odot}$ in $L_{\rm box}=150h^{-1}\mbox{Mpc}$
models may suffer from the artificial 2-body heating. Since it is
difficult to predict the possible systematic effect based on the
timescale argument alone, it is most straightforward to quantify the
effect by a careful comparison with the results in $L_{\rm
  box}=75h^{-1}\mbox{Mpc}$ runs which are almost free from the
artificial 2-body heating.
In the following analysis, we use clusters
with $M>10^{14}M_{\odot}$ and $M>10^{13}M_{\odot}$ for L150 and L75
models, respectively. Table~\ref{tab:number} indicates the number of
those clusters which satisfy the above criteria for each model at
different redshifts.

It should also be noted that one should not worry the numerical 2-body
heating at the central regions of clusters where $t_{\rm dyn} \gg
t_{\rm cool}$ even if $t_{\rm cool}\gtrsim t_{\rm 2body}$.  As noted
in \citet{Steinmetz1997}, those regions will experience a catastrophic
cooling; in fact, they are exactly the place where we attempt to
suppress the overcooling by decoupling the cold gas particles.

\subsection{Temperature -- mass relation}

A conventional analytical modeling of clusters of galaxies assumes
that the ICM is isothermal and in hydrostatic equilibrium within the
dark matter potential. Then the ICM temperature is predicted to be
\begin{eqnarray}
&& \hspace*{-1cm} k_{\rm B}T_X 
= \gamma \frac{\mu m_{\rm p}GM}{3 r_{\rm vir}} \cr
&& \sim 5.2\gamma
\left(\frac{\Omega_0\Delta_{\rm c}}{18\pi^2}\right)^{1/3} \hspace*{-0.15cm}
\left(\frac{M}{10^{15}h^{-1}M_{\odot}}\right)^{2/3}
\hspace*{-0.5cm} (1+z_{\rm f}) \,\mbox{keV} 
  \label{eq:tm-relation}
\end{eqnarray}
in terms of the cluster mass $M$, where $\Delta_{\rm c}$ is the mean
density of a virialized object at a formation redshift $z_{\rm f}$,
and $\gamma$ is a fudge factor of an order unity; if the cluster is
isothermal and its one-dimensional velocity dispersions is equal to
$\sqrt{GM/3 r_{\rm vir}}$, then $\gamma$ is the inverse of the
spectroscopic $\beta$-parameter and approximately given by $1.2$
\citep{Kitayama1997}.

\begin{figure}[tbph]
 \begin{center}
  \includegraphics[height=15cm,keepaspectratio]{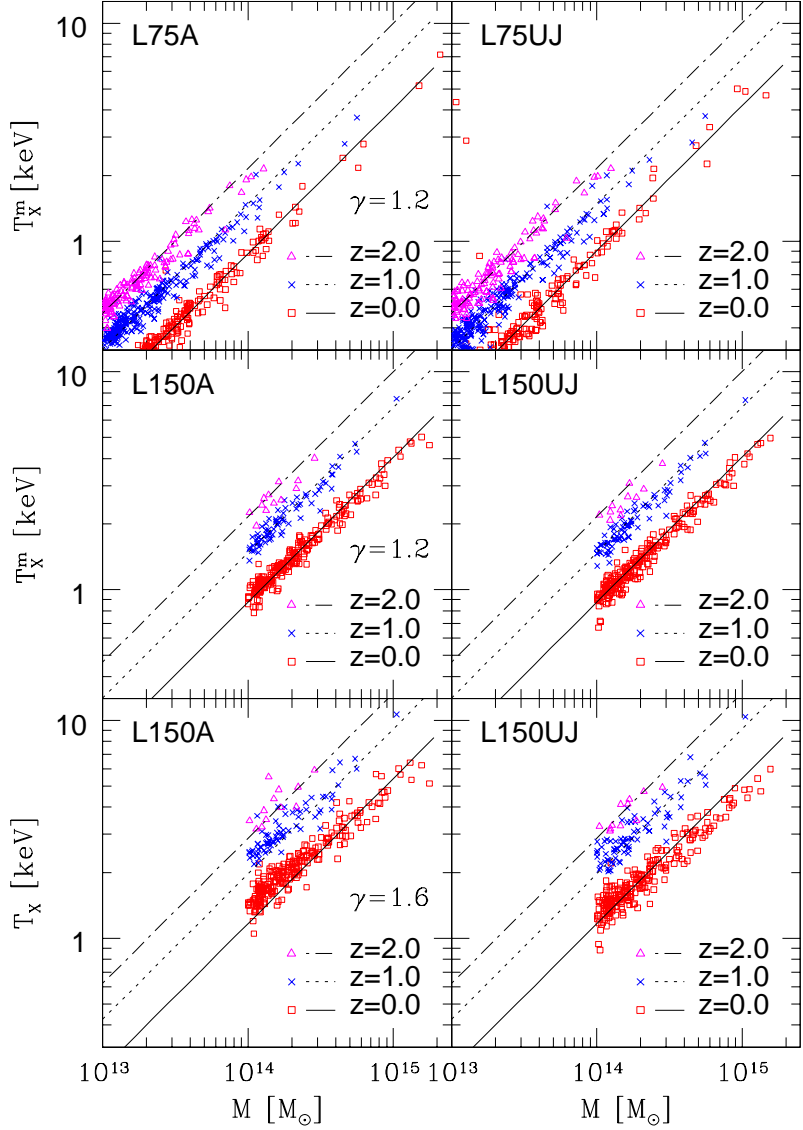}
 \end{center}
\figcaption[txm-relation.ps]{Temperature--mass relations for
  non-radiative and multiphase models at $z=0.0$, $1.0$ and $2.0$. The
  lowest panels show the emission-weighted temperature while others
  the mass-weighted temperature. Lines indicate the theoretical model
  (eq.[\ref{eq:tm-relation}]) with $\gamma=1.6$ for emission-weighted
  temperature, and $\gamma=1.2$ for mass-weighted one.
  \label{fig:tm-relation}}
\end{figure}

Since the above $T_X$--$M$ relation is the most important ingredient
in translating the Press -- Schechter {\it mass} function into the
XTF, the reliability of the conventional cluster abundance crucially
depends on the applicability of this relation.
Figure~\ref{fig:tm-relation} plots the $T_X$--$M$ relations for
non-radiative and multiphase models at $z=2$, 1 and 0. In order to
avoid the possible artificial two-body heating effect, we show the
results for clusters of $M>10^{13}\,M_{\odot}$ and
$M>10^{14}\,M_{\odot}$ in L75 and L150 models, respectively.  

This
figure implies three major conclusions; first, comparison of the upper
and middle panels indicates that the current numerical resolution is
sufficiently good and the $T_X$--$M$ relation seems to be converged.
Second, the mass weighted temperature $T_X^{\rm m}$ is well fitted to
equation (\ref{eq:tm-relation}) with $\gamma=1.2$, while the emission
weighted temperature $T_X$ is systematically higher for the same $M$.
Finally, the simulated $T_X$--$M$ relation is almost unaffected by the
phenomenological (in the current simulation) treatment of the thermal
evolution of the ICM gas, and the results of L150A and L150UJ are
almost identical. We also verify that the results of L150CJ are almost
identical to those of L150UJ implying that the presence of a UV
background does not affect our conclusions.  This is the first
successful check of the relation, made possible by the sufficiently
large volume and good resolution of our simulations.

\subsection{Luminosity -- temperature relation \label{subsec:tl}}

In contrast to the $T_X$--$M$ relation, the luminosity -- temperature
relation of X-ray clusters is not easy to predict.  This is because
the X-ray luminosity is proportional to the square of the gas density,
and thus sensitive to the thermal evolution of the ICM.  A simple
self-similar model, which predicts $L_X \propto T_X^2(1+z_{\rm
  f})^{3/2}$ \citep{Kaiser1986}, is shown to be completely
inconsistent with the observed relation of $L_X \propto
T_X^{\alpha}(1+z)^{\zeta}$ where $2.6\lesssim\alpha\lesssim 3.0$ and
$\zeta\simeq0$ \citep{Edge1991,David1993,Markevitch1998}.

\begin{figure}[tbph]
 \begin{center}
  \includegraphics[width=15cm,keepaspectratio]{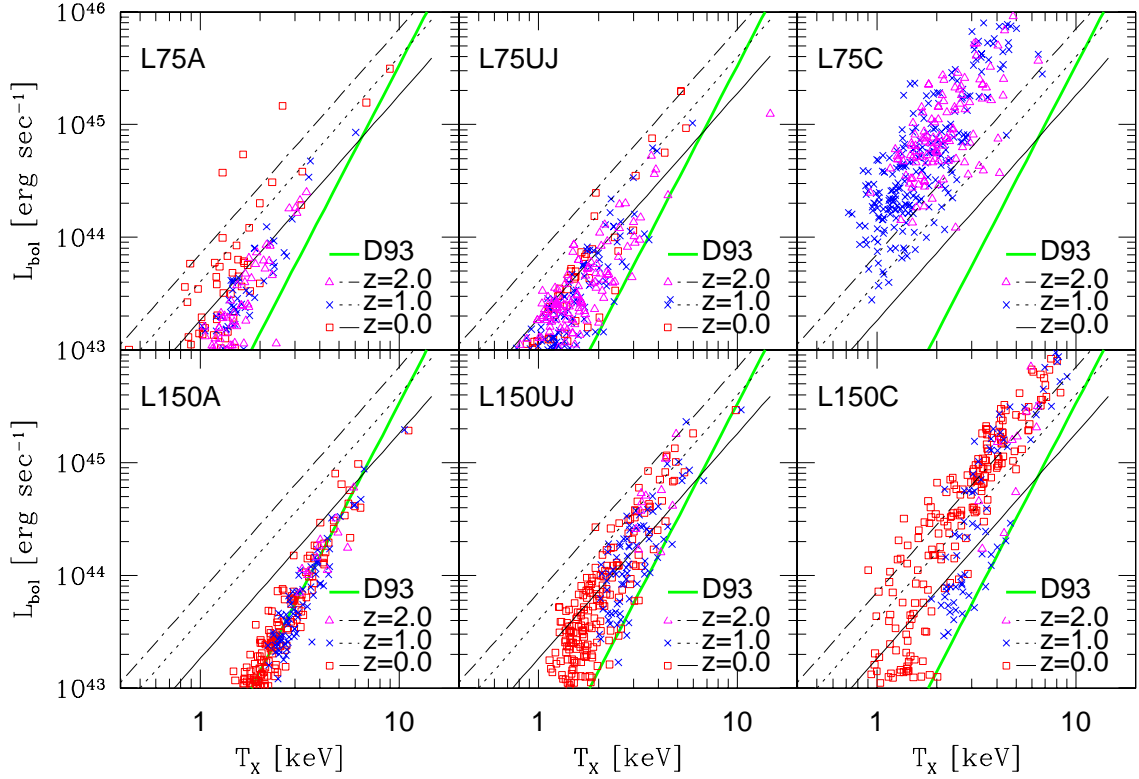}
 \end{center}
 \figcaption[ltx-relation.ps]{Luminosity--temperature relations for
   non-radiative and multiphase models at $z=0.0$, 1.0 and 2.0. Thin
   lines indicate the relations predicted from the self-similar
   evolution.  Bold lines indicate the empirical relation from
   observations (David et al. 1993).
 \label{fig:lt-relation}}
\end{figure}

The $L_X$ -- $T_X$ relation of our simulated clusters
(Fig.~\ref{fig:lt-relation}) also exhibits the difficulty to obtain a
reliable estimate of the luminosities. While it is reasonable that the
results are sensitive to the ICM thermal evolution, they are also
affected by the numerical resolution. The cluster luminosities in
$L_{\rm box}=150h^{-1}$Mpc models are systematically underestimated
relative to those in $75h^{-1}$Mpc models. Since we employ the
equal-mass particle in the simulations, the resolution problem would
be more serious for smaller clusters. Actually this explains why L150A
model produces $L_X \propto T_X^3$ accidentally although this
non-radiative model should result in $L_X \propto T_X^2$.  Thus we
conclude that the reliable estimate of the X-ray luminosities requires
much higher resolution, especially for less luminous clusters, than
ours in addition to the more physical implementation of the galaxy
formation process. Therefore, we disagree with \citet{Pearce1999b} who
concluded that the effect of cooling suppresses the X-ray
luminosities; while our results do not reach convergence either, we
find the opposite trend, i.e., the luminosities increase with the
cooling (Figs. \ref{fig:profile150} and \ref{fig:lt-relation}).  After
all, since their resolution is similar to ours, their results cannot
escape from the resolution problem, and it is premature to draw any
conclusion on the luminosities from cosmological SPH simulations.

\subsection{X-ray Temperature Function}

\begin{figure}[tbph]
 \begin{center}
  \includegraphics[width=15cm,keepaspectratio]{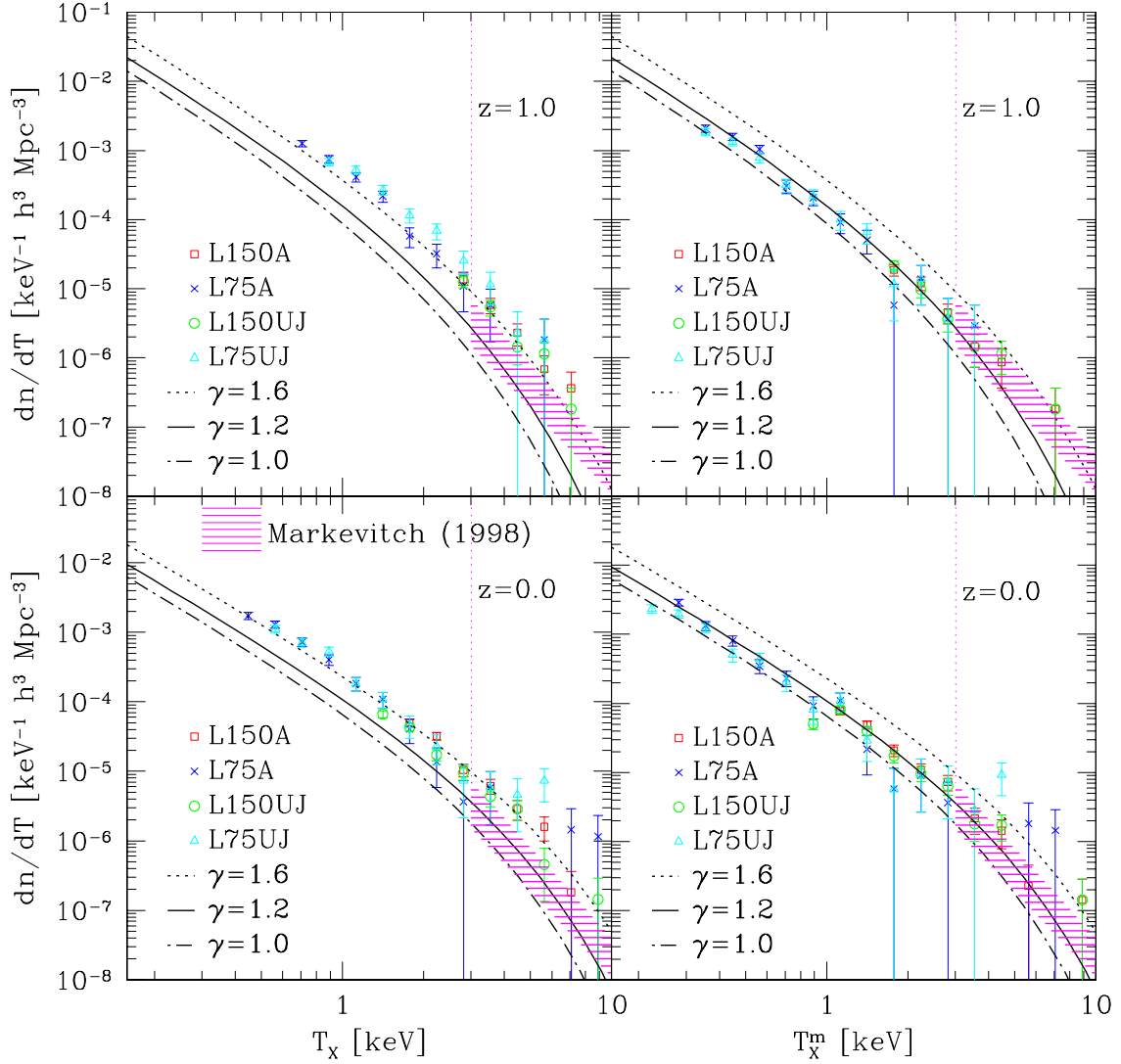} 
 \end{center}
 \figcaption[XTF.ps]{X-ray temperature functions at $z=0.0$ (lower
 panels) and $z=1.0$ (upper panels) for each model. The left panels uses
 emission weighted temperature, whereas the right ones mass weighted
 temperature. Lines are theoretical predictions using Press--Schechter
 mass function and $T$--$M$ relation (eq.[\ref{eq:tm-relation}]) with
 $\gamma=1.0$, 1.2 and 1.6. Shaded regions indicate the observed XTF of
 local ($z<0.1$) clusters by \citet{Markevitch1998}.  \label{fig:XTF}}
\end{figure}

Finally we show the XTFs in non-radiative and multiphase models at
redshift $z=0.0$ and $1.0$ (Fig.~\ref{fig:XTF}) as a function of the
emission and mass weighted temperatures. These should be compared with
the analytical prediction on the basis of the Press--Schechter mass
function \citep{Press1974} and the $T_X$--$M$ relation mentioned
above. Since we have already showed that the $T_X$--$M$ relation
agrees well with the analytical expectation, our simulated clusters in
cosmological volumes can examine the statistical reliability of the
analytical prediction of XTF for the first time.  Figure~\ref{fig:XTF}
implies that the analytical and numerical results agree well with each
other almost independently of the ICM thermal evolution model; if we
adopt $\gamma \simeq 1.6$ and $1.2$ for XTFs in terms of emission- and
mass-weighted temperatures, respectively. It should be also noted that
since the XTFs from different numerical resolutions are nearly the same
within the statistical errors, we can state that the XTFs from
our simulations do not suffer from any serious numerical artifacts.
These results justify the use of a simple analytical model for the
cluster abundance extensively applied before
\citep{White1993,Viana1996,Eke1996,Kitayama1997,Kitayama1998}. The
observed XTF of local ($z<0.1$) clusters by \citet{Markevitch1998} is
also shown in Figure~\ref{fig:XTF} and lower by factor of 2--3 than
the simulated ones at $T_{X}\simeq 3\mbox{--}9$ [keV].

\section{Conclusions}

On the basis of a series of the large cosmological SPH simulations, we
have specifically addressed the reliability of the analytical
predictions on the statistical properties of X-ray clusters. In order to
distinguish numerical artifacts from real physics, we performed
simulations with two different numerical resolutions. 
Our main findings are summarized as follows:

(i) The inclusion of radiative cooling in the high-resolution
simulations substantially change the luminosity of simulated clusters.
Without implementing the galaxy formation scheme, this leads to an
artificial overcooling catastrophe as demonstrated by
\citet{Suginohara1998}.  With a phenomenological prescription of cold
gas decoupling like \citet{Pearce1999a}, however, the overcooling is
largely suppressed. Nevertheless the predicted X-ray luminosities of
clusters are not reliable even by one order of magnitude.

(ii) In contrast to the huge uncertainties on the X-ray luminosities,
the temperature of simulated clusters is fairly robust both to the ICM
thermal evolution and to the numerical resolution, and in fact are in
good agreement with the analytic predictions commonly adopted. We also
showed that the emission-weighted temperature is 1.3 times higher than
the mass-weighted one.

(iii) The analytical predictions for the X-ray temperature function
translated from the Press -- Schechter mass function are fairly accurate
and reproduce the simulation results provided that the fudge factor
$\gamma=1.2 (1.6)$ is adopted in the mass-weighted (emission-weighted)
temperature -- mass relation (\ref{eq:tm-relation}). The XTFs simulated
in the cosmology adopted in this paper are slightly higher than the
observed one of local clusters by \citet{Markevitch1998}.

Our simulations have made statistically unbiased synthetic catalogues of
clusters of galaxies in virtue of their large simulation volume and
sufficient numerical resolution. Thus, they can be the theoretical
references which should be compared with the results from future cluster
observations with X-ray satellites including {\it XMM} and {\it
Astro-E}, and can contribute to probing cosmological parameters from
observed cluster abundances in due course. Additional radiative
processes such as heat conduction of ICM and energy feedback from
supernovae explosions, which may considerably affect the physical
properties of ICM, must be considered as future works in order to solve
the discrepancy of $L$--$T$ relation. In addition, since we consider
only one fairly specific cosmological model, it is necessary to perform
simulations under other cosmological models.

\acknowledgments

We thank an anonymous referee for pointing out the importance of the
two-body heating in the simulations. K.Y. and Y.P.J. gratefully
acknowledge the fellowship from the Japan Society for the Promotion of
Science. Numerical computations were carried out on VPP300/16R and
VX/4R at the Astronomical Data Analysis Center (ADAC) of the National
Astronomical Observatory, Japan, as well as at RESCEU (Research Center
for the Early Universe, University of Tokyo) and KEK (High Energy
Accelerator Research Organization, Japan).  This research is supported
by Grants-in-Aid by the Ministry of Education, Science, Sports and
Culture of Japan to RESCEU (07CE2002), and by the Supercomputer
Project (No.99-52) of KEK.

\clearpage

\begin{deluxetable}{lcccccc}
 \footnotesize
 \tablecaption{Summary of the simulation models\label{tab:model}}
 \tablewidth{0pt} 
 \tablehead{\colhead{model} & \colhead{$L_{\rm box}$[Mpc]} & 
 \colhead{cooling} & \colhead{$J_{21}^{0}$} & 
 \colhead{$m_{\rm gas}[M_{\odot}]$\tablenotemark{a}} & 
 \colhead{$m_{\rm dark}[M_{\odot}]$\tablenotemark{a}} &
 \colhead{cold gas decoupling}}
 \startdata 
 L150A  & $150h^{-1}$ & off & 0.0 & $2.0\times10^{10}$ &
 $1.7\times10^{11}$  & No \\
 L150C  & $150h^{-1}$ & on  & 0.0 & $2.0\times10^{10}$ &
 $1.7\times10^{11}$  & No \\
 L150UJ & $150h^{-1}$ & on  & 1.0 & $2.0\times10^{10}$ & 
 $1.7\times10^{11}$  & Yes \\
 L150CJ & $150h^{-1}$ & on  & 0.0 & $2.0\times10^{10}$ & 
 $1.7\times10^{11}$  & Yes \\
 L75A   & $75h^{-1}$  & off & 0.0 & $2.4\times10^{9}$ &
 $2.2\times10^{10}$  & No \\
 L75C   & $75h^{-1}$  & on  & 0.0 & $2.4\times10^{9}$ &
 $2.2\times10^{10}$  & No \\
 L75UJ  & $75h^{-1}$  & on  & 1.0 & $2.4\times10^{9}$ &
 $2.2\times10^{10}$  & Yes \\
 \enddata
  \tablenotetext{a}{gas and dark matter mass per particle}
\end{deluxetable}

\begin{deluxetable}{lrrr}
 \footnotesize
 \tablecaption{Number of simulated clusters that we used in the analysis\label{tab:number}}
 \tablewidth{0pt} 
 \tablehead{\colhead{model} & \colhead{$z=2.0$} & \colhead{$z=1.0$} &
 \colhead{$z=0.0$}}
 \startdata
 L150A  & 12  & 72  & 224 \\
 L150UJ & 11  & 76  & 220 \\
 L150C  & 12  & 75  & 222 \\
 L75A   & 123 & 263 & 295 \\
 L75UJ  & 117 & 241 & 272 \\
 L75C   & 120 & 245 & ---\,\,  \\
 \enddata
\end{deluxetable}

\clearpage



\clearpage

\end{document}